# Deep learning on fundus images detects glaucoma beyond the optic disc

Authors:

Ruben Hemelings[a,h]*, MS

Bart Elen[h], MS

João Barbosa Breda[a,c,d], MD PhD

Matthew B. Blaschko[e], PhD professor

Patrick De Boever[f,g,h], PhD professor

Ingeborg Stalmans[a,b], MD PhD professor

Affiliations:

[a] Research Group Ophthalmology, Department of Neurosciences, KU Leuven, Herestraat 49, 3000 Leuven, Belgium

[b] Ophthalmology Department, UZ Leuven, Herestraat 49, 3000 Leuven, Belgium

[c] Cardiovascular R&D Center, Faculty of Medicine of the University of Porto, Alameda Prof. Hernâni Monteiro, 4200-319 Porto, Portugal

[d] Department of Ophthalmology, Centro Hospitalar e Universitário São João, Alameda Prof. Hernâni Monteiro, 4200-319 Porto, Portugal

[e] ESAT-PSI, KU Leuven, Kasteelpark Arenberg 10, 3001 Leuven, Belgium

[f] Hasselt University, Agoralaan building D, 3590 Diepenbeek, Belgium

[g] University of Antwerp, Department of Biology, 2610 Wilrijk, Belgium

[h] Flemish Institute for Technological Research (VITO), Boeretang 200, 2400 Mol, Belgium

*corresponding author

Contact details

Affiliation: KU Leuven, VITO

Postal address: VITO Health, Industriezone Vlasmeer 7, 2400 Mol, Belgium

E-mail: ruben.hemelings@kuleuven.be

Financial Support: None

Conflict of Interest: No conflicting relationship exists for any author

Key words: glaucoma; fundus image; explainability; convolutional neural network; deep learning; cup-to-disc ratio

Abstract

Although unprecedented sensitivity and specificity values are reported, recent glaucoma detection deep learning models lack in decision transparency. Here, we propose a methodology that advances explainable deep learning in the field of glaucoma detection and vertical cup-disc ratio (VCDR), an important risk factor. We trained and evaluated deep learning models using fundus images that underwent a certain cropping policy. We defined the crop radius as a percentage of image size, centered on the optic nerve head (ONH), with an equidistant spaced range from 10%-60% (ONH crop policy). The inverse of the cropping mask was also applied (periphery crop policy). Trained models using original images resulted in an area under the curve (AUC) of 0.94 [95% CI: 0.92-0.96] for glaucoma detection, and a coefficient of determination ($R^2$) equal to 77% [95% CI: 0.77-0.79] for VCDR estimation. Models that were trained on images with absence of the ONH are still able to obtain significant performance (0.88 [95% CI: 0.85-0.90] AUC for glaucoma detection and 37% [95% CI: 0.35-0.40] $R^2$ score for VCDR estimation in the most extreme setup of 60% ONH crop). Our findings provide the first irrefutable evidence that deep learning can detect glaucoma from fundus image regions outside the ONH.

Glaucoma is a leading cause of irreversible blindness in our ageing society with a projected number of patients of 112 million by 2040.[1] This chronic neuropathy induces structural optic nerve fiber damage with visible changes in and outside the optic disc, ultimately leading to functional vision loss. Glaucoma is associated with characteristic changes of the optic nerve head (ONH), also called the optic disc.[2] During clinical examination and optic disc photo analysis, ophthalmologists evaluate the ONH, looking for typical changes such as generalized or focal neural rim thinning. Neuroretinal rim thinning can be quantified in fundus photos by measuring the vertical cup-to-disc ratio (VCDR).[3] The optic cup is the distinguishable excavation in the central portion of the ONH. It is typically small in normal eyes but increases with neuroretinal rim loss.[4] An elevated VCDR or interocular asymmetry > 0.2 is therefore considered suspicious for glaucoma (Figures 1B-2B).[5] Although clinicians tend to focus mainly on the optic disc for diagnosing glaucoma, retinal nerve fiber layer (RNFL) defects (adjacent to the ONH) are also known as a typical indicator of glaucomatous damage.[6] However, for the evaluation of RNFL defects, typically papillo-macular area centered red-free fundus images are used for optimal visualization of the RNFL. Even then, clinical detection of RNFL defects by red-free fundus photography is only possible after a 50% loss of the RNFL.[7]

Deep learning models and especially convolutional neural networks (CNN) are setting new benchmarks in medical image analysis. These models are finding their way in a plethora of healthcare applications including dermatologist-level classification of skin cancer[8] and identification of pneumonia on chest CT[9]. In ophthalmology, the main research focus has been the diagnostic ability of CNNs in the 'big four' eye diseases (diabetic retinopathy[10], glaucoma[11], age-related macular degeneration[12] and cataract[13]) using widely available color fundus photos and to a lesser extent optical coherence tomography (OCT) scans. Diagnostic models using deep learning can play a role in overcoming the challenge of glaucoma under-diagnosis while maintaining a limited false positive rate.[14] Successes have already been booked in the field of automated glaucoma diagnosis[11,15–31] and glaucoma-related parameters[32,33] from fundus images using CNNs. The use of end-to-end deep learning in glaucoma led to a high reference sensitivity of 97.60% at 85% specificity in a recent international challenge.[34] Unfortunately, those results came at the cost of lower insights into the decision process of the predictive model, as image features are no longer manually crafted and selected. Decision-making transparency, also referred to as explainability of the CNN, is crucial to build trust for future use of deep learning in medical diagnosis. Furthermore, it is currently unknown to what extent information outside the ONH (peripapillary area) in color fundus images is relevant to glaucoma diagnosis for deep learning. Trained deep learning models for glaucoma detection could leverage subtle changes such as RNFL thinning that human experts cannot detect. Several studies attempted to explain the deep learning model's decision in glaucoma classification from fundus images.[20,23–26,28,31,35] The majority of explainability studies[20,24,28] employed some form of occlusion[36], a technique in which parts of the test images are perturbed, and the effect on performance recorded. They mainly report significant importance of areas within the ONH. Some mentioned the presence of relevant regions directly outside the ONH in a small number of images. One major downside of occlusion testing is the violation of having a similar distribution in train and test sets. When training on a complete image, and evaluating on a perturbed image, it is impossible to assess whether the change in prediction is due to the perturbation or because the omitted information was truly (un)informative.[37] A solution is to occlude the same part of the images used for training, a principle which was recently named Remove And Retrain (ROAR).[38]

Using two pseudo-anonymized data sets of disc-centered fundus images from the University Hospitals Leuven (UZL), a large glaucoma clinic in Belgium, the goal of this work was to analyze the importance of the regions beyond the ONH and provide objective explainability in the context of glaucoma detection and VCDR estimation. To achieve this, we trained and evaluated several CNNs with a varying amount of image covered and compared performance between cover size and application (glaucoma classification/VCDR regression). We validated our glaucoma detection models on REFUGE, a public data set of 1200 glaucoma-labeled color fundus images. Our findings provide hard evidence that deep learning utilizes information outside the ONH during glaucoma detection and VCDR estimation.

Results

Table 1: Data set characteristics for vertical cup-to-disc (VCDR) regression. The upper part gives train, validation, and test splits on image, eye, and patient level. Data split was done randomly on patient level, assuring all images from the same patient to be in the same subset. The image data set consists of both glaucomatous and healthy eyes. The lower part of the table contains demographic information such as age (expressed in years) and sex (F = female, M = male), which are balanced across train, validation and test subsets. The correlation between VCDR and age is 0.29.

**Table 1 – VCDR regression (UZL)**

|  | Train | Val | Test | Total |
|---|---|---|---|---|
| Images | 16799 | 2366 | 4765 | 23930 |
| Eyes | 8587 | 1209 | 2469 | 12265 |
| Patients | 4540 | 642 | 1304 | 6486 |
| Age | | 62.9 ± 17 | | |
| Sex (F \| M) | | 0.53 \| 0.47 | | |
| VCDR | | 0.67 | | |
| Per age group: | | | | |
| 0 – 19 (% share) | | 0.53 (3%) | | |
| 20 – 39 (% share) | | 0.57 (6%) | | |
| 40 – 59 (% share) | | 0.61 (27%) | | |
| 60 – 79 (% share) | | 0.70 (49%) | | |
| ≥ 80 (% share) | | 0.77 (15%) | | |
| r (VCDR, age) | | 0.29 (P < 0.001) | | |

**Vertical Cup-to-Disc Ratio (VCDR) Regression**

A total of 23930 color fundus images (12265 eyes, 6486 individuals) were included. The UZL patient set is predominantly Caucasian, has a mean age of 63 years, and consists of a balanced sex distribution, with 53% women. The mean VCDR label is 0.67. The largest age group is 60–79 years, encompassing half of the images, with a mean VCDR of 0.7. The Pearson correlation coefficient (r) between the VCDR ground truth and age reveals a small strength of association (r=0.29, P<0.001). More demographic details on VCDR regression can be retrieved in Table 1. After a 70%/10%/20% random split on the patient level, the train, validation, and test sets contain 16799, 2366, and 4765 images, respectively.[39] The selected baseline mean absolute error (MAE) is equal to 0.19, obtained when always predicting the mean VCDR value of the 4765 test images (=0.67).

The CNN model with a ResNet[40] encoder obtained an R² value of 77% [95% CI: 0.76-0.79] between the predicted and ground truth VCDR values of the test set, translating to a very strong correlation coefficient of 0.88 [95% CI: 0.87-0.89]. On average, the VCDR predictions deviate 0.079 [95% CI:

0.077-0.081] from the ground truth VCDR, corresponding to an error reduction of 58% compared to baseline (0.19). Figure 1A plots the coefficient of determination R² as a function of the crop size for both ONH and periphery crop analysis. Evaluated crop sizes are indicated by the data markers. Kernel density estimation (KDE) plots (bottom panel of Figure 1) highlight the correlation of the bivariate distribution (ground truth versus predictions). Results are given for experiments following ONH and periphery crop policies.

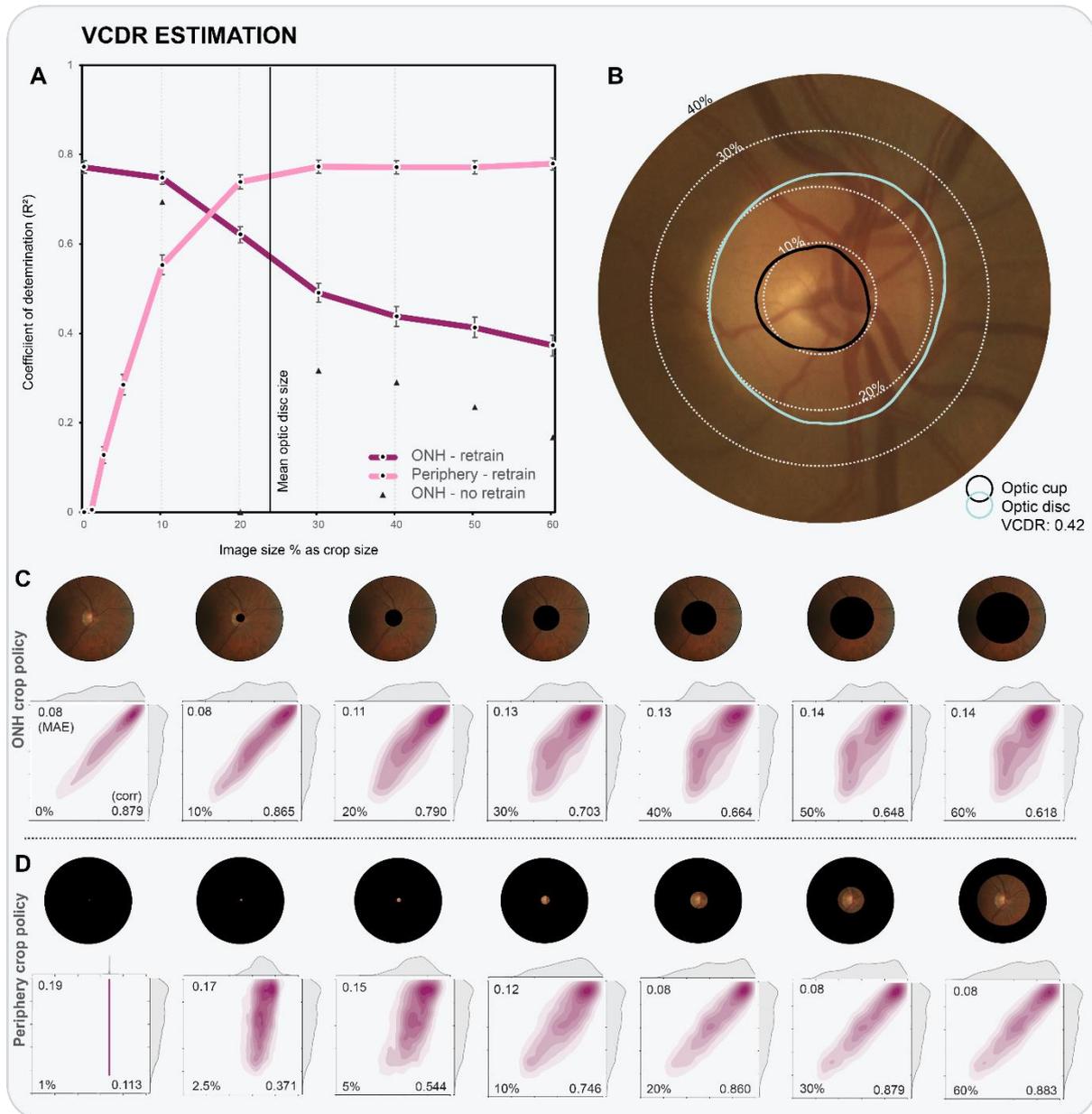

Figure 1: (A) R² values are plotted as a function of crop size, for both ONH crop and periphery crop policies. Evaluated crop sizes are indicated by data markers in the graph, and visualized in the two rows of fundus images. Results of occlusion are given as black triangles. (B) Close-up of example fundus image, with dotted lines corresponding to example crop sizes. (bottom panel) Detailed results for experiments following ONH cropping (C) and periphery cropping (D) policies. Kernel density estimation (KDE) plots with ground truth distribution on y-axis and prediction distribution on x-axis. The KDE plots also feature the MAE (top left), Pearson r (bottom right) and crop size (bottom left).

*ONH crop policy – retrain*

Experiments start at 10%, equivalent to an image that has some ONH information removed, and progress to a setting with 60%, in which the complete ONH and a large peripapillary area are removed. Examples of evaluated ONH crop policy experiments are visualized in the first row of fundus images (Figures 1C and 2C). The largest drop in performance is observed between 20% and 30%, an area that corresponds to the ONH border. With an extreme circular crop of 60% image diameter covering both ONH and a large peripapillary area, the model still explains 37% [95% CI: 0.35-0.40] of test variance, corresponding to a moderate positive correlation of 0.62 [95% CI: 0.60-0.63]. The mean absolute error is 0.142 [95% CI: 0.139-0.145], still significantly smaller than the average difference (0.19) between ground truth VCDR and mean estimated VCDR value. This finding implies that VCDR information can be retrieved from features located outside the ONH.

*ONH crop policy – no retrain (occlusion)*

The main difference with previous experiments is the lack of retraining the models with cropped images. Practically this means that the model did not encounter black circular shapes as a result of cropping during training. All evaluations (from 10% to 60%) rank significantly lower than their retraining counterpart. In the 20% setup, when the optic nerve becomes absent, the coefficient of determination even drops to zero. The model that encounters images subjected to the most extreme cropping (60%), is only able to explain 17% of the test variance (versus 37% with retraining).

*Periphery crop policy - retrain*

These experiments invert the previous crop policy, with an increasing amount of ONH and periphery (from 20% onwards) being accessible to the CNN (see fundus images in Figures 1D and 2D). When we cropped the images so that only 1% of the image size is visible, we observe no significant results ($R^2$=0.01 [95% CI: 0.00-0.01], MAE=0.189 [95% CI: 0.186-0.193]) and weak correlation (r=0.11 [95% CI: 0.09-0.14]). Already with 2.5% of the fundus image visible, the model yields predictions that outperform baseline for VCDR estimation ($R^2$=0.13 [95% CI: 0.11-0.15]; MAE=0.174 [95% CI: 0.170-0.177], Pearson r=0.37 [95% CI: 0.35-0.40]). A setup with 30% of the image radius used, and therefore with fully visible ONH, obtains results as high as a setup with a complete image ($R^2$=0.77 [95% CI: 0.76-0.79]; MAE=0.0784 [95% CI: 0.0764-0.0806]). Hence, the CNN only requires the intact ONH to estimate the VCDR as accurately as a CNN trained using original 30° disc-centered images.

Table 2: Data set characteristics for glaucoma classification. Equivalently to Table 1, the upper part gives train, validation, and test splits on image, eye, and patient level, while the lower half yields demographic information. The data set contains a balanced set of 55% glaucoma (G) and 45% non-glaucoma (G̶) images. The bivariate correlation between having glaucoma and age is moderate (0.56).

**Table 2 – Glaucoma detection (UZL)**

|  | Train | Val | Test | Total |
|---|---|---|---|---|
| Images | 9541 | 1368 | 2643 | 13551 |
| Eyes | 4904 | 698 | 1407 | 7009 |
| Patients | 2514 | 355 | 723 | 3592 |
|  |  |  |  |  |
| Age | | 59.5 ± 21 | | |
| Sex (F \| M) | | 0.50 \| 0.50 | | |
| Glaucoma (G \| G̶) | | 0.55 \| 0.45 | | |
| Per age group: | | | | |
| 0 – 19 (% share) | | 0 \| 1 (8%) | | |
| 20 – 39 (% share) | | 0.03 \| 0.97 (9%) | | |
| 40 – 59 (% share) | | 0.44 \| 0.55 (22%) | | |
| 60 – 79 (% share) | | 0.71 \| 0.29 (45%) | | |
| ≥ 80 (% share) | | 0.82 \| 0.18 (15%) | | |
| r (Glaucoma, age) | | 0.56 ($P < 0.001$) | | |

**Glaucoma Classification**

For the glaucoma classification problem, a selection of 13551 images (7009 eyes, 3592 individuals) was withheld based on the procedure outlined in Methods. Population sample characteristics include a mean age of 60 years and a 50% sex distribution. A 70/10/20 data split was used for defining the train, validation and test set of 9541, 1368, and 2643 images, respectively. A little over half (55%) of the patients had a glaucoma diagnosis. In this task, no glaucoma patients were found in the youngest age category. We observed a moderate point biserial correlation (p=0.56, P<0.001) between glaucoma and age in this patient set. Additional information can be retrieved from Table 2.

Our glaucoma classification CNN obtains a benchmark AUC of 0.94 [95% CI: 0.92-0.96] on the UZL test set of 2643 images. Similarly to the graph of Figure 1A, we plot the AUC values as a function of crop size for ONH crop (with/without retrain) and periphery crop policies in Figure 2A.

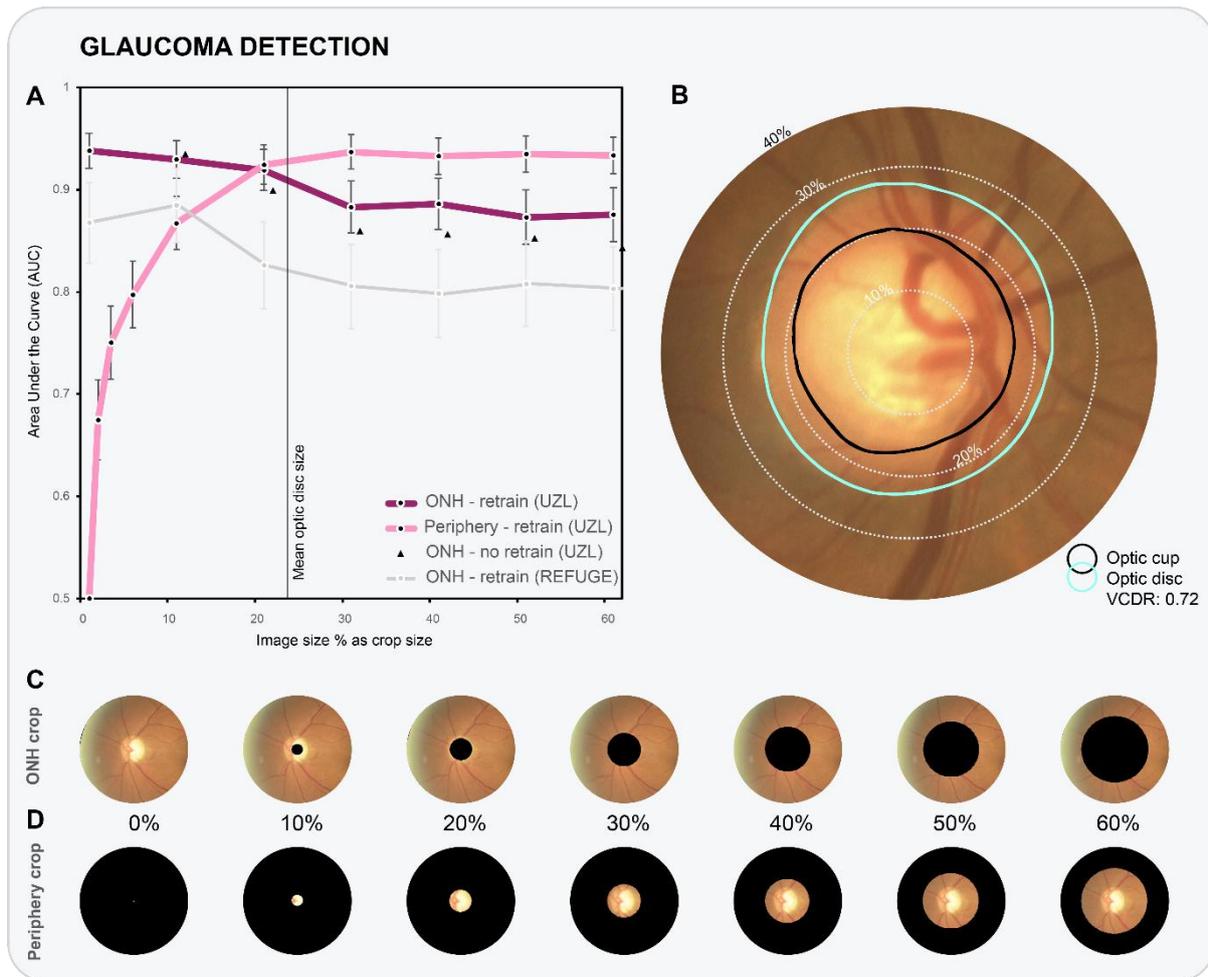

Figure 2: (A) AUC values are plotted as a function of crop size, for both ONH crop and periphery crop policies. Evaluated crop sizes are indicated by data markers in the graph. (B) Close-up of example fundus image (UZL), with dotted lines corresponding to example crop sizes. (bottom panel) Two rows of processed fundus images of REFUGE data, with both ONH cropping (C) and periphery cropping (D) policies applied.

*ONH crop policy – retrain*

The performance of glaucoma classification remains comparable until 20% ONH crop policy, after which a significant decrease of four percentage points is observed (30% ONH crop: AUC=0.88 [95% CI: 0.86-0.91]). Remarkably, there is no statistical difference in performance between models with ONH crop at 30-60%. The model is still performant with 60% of image radius covered (AUC=0.87 [95% CI: 0.85-0.88]).

*ONH crop policy – no retrain (occlusion)*

The impact of standard occlusion is smaller on AUC values than it was for $R^2$ values in VCDR regression. The largest difference on AUC values is recorded at 40% crop, with a gap of 0.03.

*ONH crop policy – REFUGE*

The CNNs trained on UZL data generalize well to unseen REFUGE data. The glaucoma classification model trained using original fundus images obtained an AUC of 0.87 [95% CI: 0.83-0.91] on the 1200 external fundus images. Similar to ONH cropping on UZL data, classifier performance remained stable from 30% ONH cropping onwards, still reporting an AUC of 0.80 [95% CI: 0.76-0.84] at 60%.

*Periphery crop policy - retrain*

With an extreme cropping from the periphery with only 1% of the fundus image visible a significant glaucoma classification with an AUC of 0.67 [95% CI: 0.64-0.71] is achieved. As from 20% cropping, the 95% AUC confidence intervals between crop policies overlap. Additional image information did not lead to significantly better performance (60% periphery crop: AUC = 0.93 [95% CI: 0.92-0.95]). The gap between 10 and 20% crop size in obtained AUC values confirms the influence of the optic disc border on glaucoma detection performance.

**Visualizing salient information**

Saliency maps are aligned on ONH, mirrored horizontally for left eyes, and averaged on all test images (4765 and 2643 for VCDR and glaucoma detection, respectively) to detect recurrent patterns. In the experiment trained with complete images (first column), we observe a diffuse saliency map for glaucoma detection, weakly highlighting superior (upper) and inferior (lower) regions within and adjacent to the ONH. For VCDR regression, saliency is visibly more concentrated in both supero- and inferotemporal sectors within the disc. With progressive covering of the optic disc information, saliency maps start to highlight more supero- and inferotemporal sectors outside the ONH (see last column in Figure 3). These sectors are characterized by the presence of the thickest RNFL in the retina. In the extreme case of 60%, saliency concentration is similar for glaucoma classification and VCDR regression.

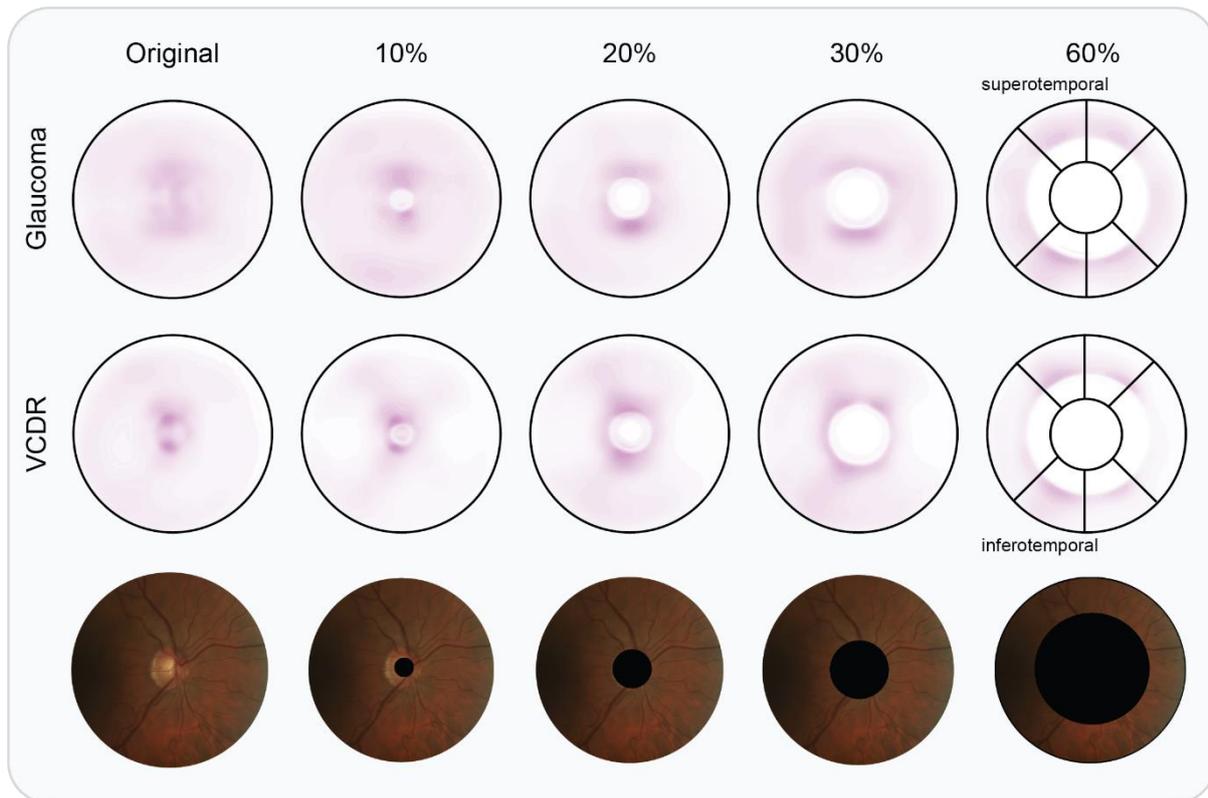

Figure 3: Averaged saliency maps for a selected number of ONH crop experiments. Top row: glaucoma detection saliency, averaged over 2643 UZL test images; middle row: VCDR regression saliency averaged over 4765 UZL test images; bottom row: a complimentary UZL fundus image of a right eye for illustrative purposes. In the last column, we draw the sectors used by glaucoma experts to locate damage. Infero- and superotemporal (bottom- and top-left, respectively) sectors are the locations most commonly damaged by glaucoma.

Discussion

We are the first to combine accurate VCDR estimation and glaucoma detection in one study using end-to-end deep learning. On top of that, we analyze the importance of ONH and peripapillary regions in deep learning experiments in which systematically cropped fundus images were used. We revealed the presence of significant pixel information on glaucoma and VCDR outside the ONH, reporting AUC values up to 0.88 and an R² score of 37% in the most extreme setup, respectively. Finally, we also uncover recurrent patterns in averaged saliency maps, pointing at locations with the thickest RNFL, that are important for the performance of the deep learning models.

The most striking observation is that significant performance in glaucoma detection and VCDR estimation can be achieved without access to the ONH. This is relevant because it answers the clinical question whether significant glaucomatous features are present outside the ONH in fundus images, even if there are no visible localized RNFL defects. Both clinicians and automated screening software can therefore focus on the peripapillary area in eyes that suffer from conditions that hamper ONH (neuroretinal rim) assessment. One such example is (pathological) myopia, in which the induced skew/tilt of the optic cup leads to decreased contrast between neuroretinal rim and optic cup.[41] Furthermore, pathological myopia was found to be a major culprit of false positives and negatives in glaucoma detection using deep learning.[11,24] Further research is needed to assess the discriminatory power of peripapillary regions for glaucoma detection in myopes.

Particularly in case of the VCDR, a parameter that is directly derived from the ONH, one would anticipate trivial performance if no ONH information is available. Yet, our trained ResNet-50 model is still able to explain over 37% of the variance in the test data, in the extreme setting of 60% crop. Hence, a large amount of information is being retrieved from correlated image characteristics in the peripapillary area of the fundus image. Most notably, a strong inverse correlation exists between VCDR and average RNFL thickness.[42] In related work, Medeiros et al demonstrated that average RNFL thickness values can be deducted from optic disc-centered fundus images.[32] This raises the question whether their trained model leverages ONH information, or whether it is truly focusing on RNFL features. Their activation maps suggest that both ONH and adjacent RNFL are the most informative. In our study, we confirm that there is information on VCDR estimation and glaucoma detection outside the ONH. The average saliency map for VCDR (Figure 3, 2nd row) indicates a pattern in infero- and superotemporal regions, which matches with RNFL locations prone to glaucoma-induced damage.

After a drop between 20-30% crop, performance in glaucoma classification is stable at AUC values around 0.88. Both the drop in performance and the subsequent stabilization are insightful. First, the drop between 20-30% (0.92-0.88 AUC) indicates that the most unique information is situated at the border and or adjacent region of the ONH. The unique information relevant to glaucoma detection at the border of the ONH could be not be represented by other regions of the image. The ONH border corresponds with the location of the thickest RNFL in human eyes and area where cupping is clearly visible.[43] Next to that, glaucoma-related information is omnipresent in the peripapillary area. Unlike VCDR, which sees its performance drop progressively as crop size increases (Figure 1A), glaucoma detection manages to maintain stable performance in two independent test sets (Figure 2A). The most likely explanation is that the CNN detects subtle changes in reflectance characteristics of the RNFL distant from the ONH. This hypothesis is strengthened by the concentrated saliency in locations with thickest RNFL (see top panel of Figure 3).

End-to-end VCDR regression using deep learning has not been reported prior to this work. Previous reports approach VCDR prediction as a segmentation problem to obtain contours of both optic disc and cup using deep learning.[34] Another recent study used fundus images to regress the Bruch's membrane opening minimum rim width (BMO-MRW), a novel parameter to quantify neuroretinal rim tissue, reporting similar performance metrics as in our study ($R^2$=0.77 in both studies).[33]

We achieve expert-level performance for glaucoma detection on two independent test sets (AUC=0.94 and AUC=0.87 for UZL and REFUGE data, respectively). Related work differs in the definition of ground truth of glaucoma. On one hand, some authors use the clinical diagnosis based on a set of exams and modalities (including our study), reporting AUC values between 0.900 and 0.995.[19,22,23,31,34] On the other hand, other authors define referable glaucoma exclusively based on (mostly) objective image features that indicate damage to the optic nerve, present in or outside the disc (AUC 0.863-0.996).[17,20,24–26,28–30,35] Our multimodal glaucoma diagnosis ground truth is regarded as superior by glaucoma experts as we teach the model to look for signs of glaucoma that have potentially been detected on a different modality than fundus images, such as diffuse RNFL loss identified with OCT[44].

The majority of glaucoma detection studies featuring CNN explainability use some type of occlusion testing, a technique in which part of the input image is perturbed during test-time, with the change in output probability recorded. This is different from our methodology, as we also modify the training images in a standardized manner (see bottom panel of Figure 1 for examples with various crops). For completeness, we also reported results without retraining (black triangles in Figures 1A and 2A). One can observe a consistent performance drop for all crop sizes. Models were forced to look for other information outside the optic disc, as it was removed during training. For VCDR regression without retraining, the model becomes insignificant at 20%. This is likely due to the importance of the ONH border, which is obscured at this level of occlusion. More concretely, Liu et al.[24] provide a single occlusion map for an example with clear RNFL defect, highlighting the latter and the complete optic disc. Li et al.[28] trained and tested on ONH-cropped fundus images and provide an occlusion map for a glaucoma case and non-glaucoma case. They identify the neuroretinal rim as the most informative area. Christopher et al.[20] provide mean occlusion maps on ONH-cropped images, highlighting superior and inferior regions of the ONH border.

Other widely used techniques that produce saliency maps have also been applied in a glaucoma detection context. Li et al.[25] used guided backpropagation, a signal estimator that visualizes input patterns responsible for neuron activation in deeper layers. They provide six examples that highlight superior and inferior rim area, localized RNFL defects, or the complete ONH. They also provide a class activation map (CAM) for the same examples, which identified larger regions of interests. Phan et al.[23] also employ CAM, and display two heatmaps, both highlighting the complete ONH. They also trained on a ONH crop, equivalent to our 30% periphery crop policy, and achieved comparable results as with the larger field of view. Keel et al.[35] noticed saliency within the ONH for 90% of glaucoma cases using a traditional sliding-window approach, with the remaining 10% highlighting the inferior/superior RNFL arcades and nontraditional ONH areas (central cup or nasal or temporal neuroretinal rim). Our group[31] conducted saliency analysis based on gradients[45]. Through the averaging of 30 ONH-aligned glaucomatous images, this study revealed a recurrent pattern in inferotemporal and superotemporal regions neighboring the ONH, as well as the ONH itself. In the current study, we improve on this method by averaging over 4765 and 2643 test images for VCDR regression and glaucoma detection, respectively. On top of that, we visualize the effect of standardized cropping on the saliency maps. The latter confirm the presence of salient information in supero- and inferotemporal regions outside the ONH.

Previous studies have discovered that a large amount of information related to ageing is present in fundus images.[46,47] It is therefore relevant to assess the influence of age in our trained models. In the data used for our study, we observed a low correlation between age and VCDR, and a moderate correlation between age and glaucoma. As model performance is more elevated than the observed correlation, we hypothesize that the model could be using ageing information to a limited degree.

The main strengths of our work are threefold. First, we report explainability analysis on two large labeled test sets for both glaucoma detection and VCDR estimation, adhering to the best practices for healthcare-related deep learning manuscripts.[48] Through the comparison of the two related applications, we observe that information on glaucoma can be retrieved in the peripapillary area for both glaucoma detection and VCDR estimation, but more so for the former. We improve on the occlusion technique described in the majority of glaucoma detection papers by employing a standardized cropping methodology on both train and test images. Finally, we provide additional insights through the systematic averaging of ONH-aligned saliency maps obtained on all test images.

Our research also suffers from several limitations. As we applied masks of fixed size in our experiments, which might lead to a small variation in visible features across fundus images due to variation in ONH size across the study population. We motivate the choice for a fixed size by the observation that ONH size and shape could introduce a potential bias in glaucoma diagnosis or VCDR estimation. Because the ONH border is an important image region for glaucoma detection and VCDR estimation, we envisage a future study in which we define the crop radii relative to the optic disc size to assess what exact features are salient at the ONH border. Furthermore, we did not explicitly assess the influence of myopic changes (e.g. PPA, titled discs) in our experiments, as this information was not available. To counter potential biases of disc size and myopia-induced disc changes, we included experiments with large cropping (>40%) that support our findings. In the future, we will incorporate explicit myopia information in our study design. We did not analyze the role of the disease stage in our experimental setup. Future experiments could investigate whether deep learning explainability analysis matches with the long-standing theory that glaucoma damage typically starts with infero-temporal and supero-temporal damage.[49] We opted for the omission of suspected glaucoma cases to have a clear idea on CNN explainability in certain cases. It would be of interest to extend our approach to glaucoma suspects in future work.

In conclusion, we present a sound methodology that conclusively supports that deep learning can reliably identify glaucoma-induced damage outside the ONH. We advance upon current explainability methods in glaucoma modelling using CNNs, often presented as a secondary objective in glaucoma detection papers. Our findings indicate that detection of glaucoma using deep learning is also possible for individuals which are difficult to judge in clinical fundoscopy because of deviating anatomy of the ONH. This is highly promising for the broader clinical applicability of deep learning in computer-aided glaucoma screening and follow-up.

Methods

**Datasets**

UZL

The retrospective cross-sectional dataset of 37627 stereoscopic color fundus images and corresponding meta-information of 9965 pseudonymized patients were extracted from the glaucoma clinic records of the University Hospitals Leuven (UZL), Belgium. The discrepancy between the number of images and patients is due to the inclusion of both eyes and the stereoscopic nature of the images, generating two images per eye. Patients imaged at this glaucoma clinic are likely to have or to be at risk of acquiring glaucoma, since it is a tertiary reference center for this pathology. Hence, the number of healthy eyes in the dataset is low. This work is part of the larger study on "automated glaucoma detection with deep learning" (study number S60649), approved by the Ethics Committee Research UZ/KU Leuven in November 2017. All steps in the study design adhered to the tenets of the Declaration of Helsinki. Informed consent was not required due to the retrospective nature and waived by the Ethics Committee Research UZ/KU Leuven. Patient reidentification is not possible as the link between patient ID and study ID was deliberately removed. Selected fundus images displayed in this manuscript comply with informed consent or belong to the public domain.

The raw export included relevant information such as demographics, results from examinations at the glaucoma clinic (e.g. intra-ocular pressure measurements, disc assessment through fundoscopy, retinal nerve fiber layer thickness measurements, amongst other), and diagnoses in the form of ICD-9 codes, next to the stereoscopic set of color fundus images. Diagnoses were established based on presence of characteristic ONH damage and corresponding functional visual field loss, as well as metadata such as intra-ocular pressure, family history, central corneal thickness and other known risk factors for glaucoma. All fundus images were captured with a Zeiss VISUCAM® (Carl Zeiss Meditec, Inc., Dublin, CA), set at a viewing angle of 30°, with the optic disc centered as is common in a glaucoma follow-up context. Data export was limited to the most recent consultation per patient.

REFUGE

The first edition of Retinal Fundus Glaucoma Challenge (REFUGE) was held as part of the Ophthalmic Medical Image Analysis (OMIA) workshop at MICCAI 2018, representing the first initiative to provide a unified evaluation framework for glaucoma detection from fundus images.[34] The data set consists of 1200 fundus images, imaged using two different types of fundus cameras (Zeiss VISUCAM and Canon CR-2) at a 45° viewing angle. 10% of the data was labeled as glaucomatous (120 images). We processed (cropped) the images to obtain a 30° viewing angle, in order to have similarly-looking images as the UZL data.

**Sample selection (UZL)**

During a typical fundoscopic examination at the University Hospitals Leuven, the supervising glaucoma specialist would visually inspect the optic disc and assess the VCDR. A drawback of this label is its subjectivity and granularity: estimates are rounded at 0.05, ranging from 0 to 1. Then again, a real-time assessment allows for better interpretation of the true cupping due to the expert's repositioning ability and true depth awareness. A total of 23930 color fundus images (12265 eyes, 6486 individuals) were found to have a valid VCDR label assigned by a glaucoma specialist at the time of examination. This represents 65% of the raw export.

Selected glaucoma patients were limited to patients with a glaucoma ICD-9 code related to primary open-angle glaucoma (POAG) or normal-tension glaucoma (NTG) assigned to both eyes (or single eye in case of a monophthalmic patient). Non-glaucoma patients were selected based on the absence of a glaucoma-related ICD-9 code. The non-glaucoma group could still include other optic neuropathies and comorbidities. A total of 13551 images (7009 eyes, 3592 individuals) were selected for the detection task, representing 36% of the raw export. A large group of the unused images belong to patients with a glaucoma suspect or other type of glaucoma diagnosis as ICD code. 7455 images were present in both VCDR and glaucoma subsets, representing 31% and 55% of the subsets, respectively.

Tables 1 and 2 describe demographic and clinical characteristics of the patient sets for VCDR regression and glaucoma detection.

**Optic disc localization & image quality control**

In order to crop the optic disc, or surrounding periphery, a state-of-the-art disc localization tool is warranted. We trained a U-Net[50] using five publicly available datasets for optic disc segmentation.[34,51–54] Our proprietary generalizable deep learning algorithm for optic disc detection found a disc in 98.2% of the UZL images for vertical cup-to-disc ratio regression and glaucoma detection. The remaining 1.8% in which no disc was detected were discarded from further analysis. On average, the vertical optic disc diameter represents 23% of the total amount of pixels per photo in this data set of 30° color fundus images. The REFUGE data set features publicly available labels for optic disc segmentation, which facilitated optic disc localization.

**Modelling**

A single deep learning framework was employed across the two tasks and experiments to allow for proper comparison. ResNet, introduced in 2015, is a deep learning network that was the first to employ residual or skip connections, which sped up training of deep models.[40] These residual connections skip one or more layers to allow gradients to flow through the network directly, omitting the passage through non-linear activation functions. We selected the standard version which consists of 50 layers, pretrained on ImageNet[55], and made it fully-convolutional by replacing the fully-connected layers with a global average pooling operation. All network layers were unfrozen, as there are sufficient data points to train the complete layer set. All network settings are equal for both VCDR regression and glaucoma detection, except for the final activation layer and loss function. Regression tasks require a linear activation at the output neuron, and are optimized through the mean squared error loss (MSE), whereas binary classification is achieved through a sigmoid activation and binary cross-entropy loss.[56] Adam[57] was used as optimizer with a base learning rate of 0.0001. The latter was reduced through the multiplication of factor 0.75 after ten successive epochs of no observed reduction in validation loss. The final models expect a preprocessed 512 x 512 color fundus image as input, and a (semi-)continuous value between 0 and 1 (VCDR regression) or binary variable (glaucoma detection) as target variable. Model development was done in Keras[58] v2.2.4 with TensorFlow[59] v1.4.1 backend.

Color fundus images were subject to a series of preprocessing steps prior to model input. First, the images were cropped to a square shape (resolution of 1444 x 1444 pixels). Subsequently, a widely-used local contrast enhancement through background subtraction was used to correct uneven illumination.[60] Due to the latter, the border of the region of interest (ROI) is overexposed. To counter this, we apply a black circular clipping mask that has a radius of 10 pixels smaller than the ROI. Next, the optic disc or periphery region is cropped, depending on the experiment type (referred to as ONH crop and periphery crop). The crop consists of a blacked out region outside (respectively inside) a circular region centered at the centroid of the discovered optic disc with a diameter defined as a percentage of image width (see Figures 1-3). Crop diameters run from 0 (original image) to 60% of image width. ONH crop and periphery experiments resulted in seven and nine models per application, respectively. Given the lower number of test images in the glaucoma detection application, experiments were repeated three times and results averaged to obtain robust results. As a results, a total of 64 models were trained and evaluated. Finally, the images are resized to the aforementioned 512 x 512 expected model input, with pixel values rescaled between 0 and 1. Data augmentation techniques included horizontal mirroring, elastic deformation, brightness shift, and cutout[61].

**Evaluation**

For VCDR regression, we use the mean absolute error (MAE), the coefficient of determination ($R^2$), and the Pearson correlation coefficient (r) to allow quantitative comparison between experiments. The MAE can be interpreted as the expected error in VCDR prediction, and compared with the baseline error of the test set (=0.19). The latter is defined as the average error term when predicting the mean value of the test set. We make use of kernel density estimation to provide a visual indicator of goodness of fit across experiments.

Glaucoma detection is evaluated using the area under the receiver operating characteristic curve (AUC). Results are reported on patient level, as glaucoma screening is typically done per patient (patient is referred or not). The conversion to prediction on patient level consists of taking the maximum value of predictions for all images per patient, mimicking the clinical diagnosis in which a patient is diagnosed with glaucoma in case one of the eyes features signs of glaucoma onset.

For both tasks, we plot the main evaluation metric ($R^2$ and AUC) as a function of crop size, along with the performance for a setting without retraining. 95% confidence intervals (CI) for all metrics were obtained using bootstrap (5000 iterations). Final results for glaucoma detection were obtained through the averaging of three runs per experiment to ensure more accurate results (test sets are smaller than for VCDR regression). Individual saliency maps based on gradients are generated using iNNvestigate.[62] They are subsequently realigned using the centroid of the segmented optic disc, and averaged to a single heat map. Saliency maps of left eyes are mirrored horizontally, to guarantee consistency in the averaging process.

## Acknowledgments

The first author is jointly supported by the Research Group Ophthalmology, KU Leuven and VITO NV. This research received funding from the Flemish Government under the "Onderzoeksprogramma Artificiële Intelligentie (AI) Vlaanderen" programme. No outside entities have been involved in the study design, in the collection, analysis and interpretation of data, in the writing of the manuscript, nor in the decision to submit the manuscript for publication. Thus, the authors declare that there are no conflicts of interest in this work.